\documentstyle[aps,prl,epsf,twocolumn,floats]{revtex}

\begin{document}
\draft

\title{Superconductor-proximity effect\\in chaotic and integrable billiards}
\author{J. A. Melsen, P. W. Brouwer, K. M. Frahm, and C. W. J. Beenakker}
\address{Instituut-Lorentz, University of Leiden\\
P.O. Box 9506, 2300 RA Leiden, The Netherlands}
\date{To be published in Physica Scripta}
\maketitle

\narrowtext
\begin{abstract}
We explore the effects of the proximity to a
superconductor on the level density of a billiard for the two extreme
cases that the classical motion in the billiard is chaotic or integrable.
In zero magnetic field and for a uniform phase in the superconductor, a
chaotic billiard has an excitation gap equal to the Thouless energy.
In contrast, an integrable (rectangular or circular) billiard has a
reduced density of states near the Fermi level, but no gap. We present
numerical calculations for both cases in support of our analytical results.
For the chaotic case, we calculate how the gap closes as a function of
magnetic field or phase difference.
\end{abstract}
\bigskip


There exists a way in quantum mechanics to distinguish classically chaotic
systems from integrable systems, by looking at correlations between energy
levels \cite{Bohigas,Berry}.
In a billiard with integrable dynamics, on the one hand, the
energy levels are uncorrelated, and the spectrum has Poisson statistics.
In a chaotic billiard, on the other hand, level repulsion leads to strong
correlations and to Wigner-Dyson statistics \cite{Mehta}.
Although the level correlations are different, the mean level density does
not distinguish between chaotic and integrable billiards.

In a recent paper \cite{Melsenetal} we have shown that the proximity to a
superconductor makes it possible to distinguish chaotic from integrable
billiards by looking at the density of states. In a chaotic billiard, on
the one hand, we have found (using random-matrix theory) that the coupling
to a superconductor by means of a point contact opens a gap in the
density of states of the order of the Thouless energy
$E_{\rm T}=N\Gamma \delta/2\pi$. Here
$N$ is the number of transverse modes in the point contact, $\Gamma$ is the
tunnel probability per mode, and $\delta$ is
half the mean level spacing of the isolated billiard. In an integrable
rectangular billiard, on the other hand, the density of states vanishes
linearly near the Fermi level, without a gap. We have argued (using the
Bohr-Sommerfeld approximation) that the absence of an excitation gap is
generic for integrable systems. In these Proceedings we present
numerical calculations for a chaotic billiard in support of the random-matrix
theory, and we consider the effects of a magnetic field perpendicular to
the billiard.
We also present calculations for an integrable circular billiard, both
exact and in the Bohr-Sommerfeld approximation.

\begin{figure}
\epsfxsize=0.95\hsize
\epsffile{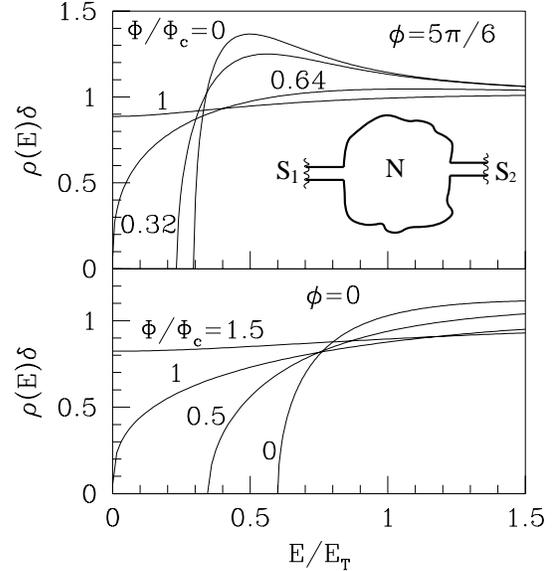}
\caption{\label{fig:fig1}Density of states of a chaotic billiard
coupled to two superconductors by identical ballistic point contacts, for
four values of the magnetic flux $\Phi$ through the billiard. The phase
difference $\phi$ between the superconductors equals $5\pi/6$ and $0$ in
the top and bottom panel, respectively.
The curves are computed from Eqs.\ (\ref{eq:resolvent}) and
(\ref{eq:dysongamma}). The Thouless energy is given by
$E_{\rm T}=N \delta/2\pi$, and
the critical flux $\Phi_{\rm c}$ is defined by Eqs.\
(\ref{eq:crossoverpar}) and (\ref{eq:critical}).
}
\end{figure}

The system studied is shown schematically in the inset of Fig.\ \ref{fig:fig1}.
A billiard consisting of a normal metal (N)
in a perpendicular magnetic field $B$ is connected to two
superconductors (S$_1$, S$_2$) by narrow leads, each containing
$N/2$ transverse modes at the Fermi energy $E_{\rm F}$. The order parameter
in S$_1$ and S$_2$ has a phase difference $\phi\in [0,\pi]$.
Mode $n$ couples to a superconductor with phase $\phi_n=\phi/2$ for $1 \leq n
\leq N/2$, $\phi_n =-\phi/2$ for $1+N/2 \leq n \leq N$.
For simplicity, we assume in this paper that there is no tunnel barrier in
the leads ($\Gamma=1$). (The generalization to $\Gamma \not=1$ is
straightforward \cite{Melsenetal}.)

The excitation spectrum of the billiard is discrete for energies
$0 < E < \Delta$, where $\Delta$ denotes the excitation gap in the bulk
of the superconductors.
We compute the spectrum starting from the scattering formulation of Ref.\
\onlinecite{Beenakker91}. The billiard with the leads but without the
superconductors has an $N\times N$ scattering matrix $S_0(E)$, at an energy
$E$ relative to the Fermi level. The modes in the leads which form the basis
of $S_0$ are chosen such that their wave functions are real at the NS
interface. Evanescent modes in the leads are disregarded.
The energy dependence of $S_0$ is set by the
Thouless energy $E_{\rm T}$, which is inversely proportional to the dwell
time of an electron in the billiard. Andreev reflection at the superconductors
scatters electrons (at energy $E > 0$) into holes (at energy $-E$), with a
phase increment $-\phi_n - \arccos(E/\Delta)$. (Normal reflection at the
superconductors can be neglected if $\Delta \ll E_{\rm F}$.) We assume that
$E$ and $E_{\rm T}$ are both $\ll \Delta$, so that we may replace
$\arccos(E/\Delta)$ by $\pi/2$, while retaining the $E$-dependence of $S_0$.
The excitation spectrum is obtained from the determinantal equation
\cite{Beenakker91}
\begin{equation}
\mbox{Det}[1+S_0(E) e^{i\mbox{\boldmath $\scriptstyle \phi$}}S^*_0(-E)
e^{-i \mbox{\boldmath $\scriptstyle \phi$}}]=0,
\label{eq:resdet}
\end{equation}
where {\boldmath $\phi$} is a diagonal matrix with the phases
$\phi_n$ on the diagonal.

The scattering matrix $S_0$ can be expressed in terms of
the Hamiltonian matrix $H$ of the isolated billiard
\cite{MahauxWeidenmueller},
\begin{equation}
S_0(E) = 1-2\pi i W^{\dagger}(E-H+i \pi W W^{\dagger})^{-1}W.
\label{eq:det}
\end{equation}
The dimension of $H$ is $M\times M$, and the limit $M\to \infty$ will be
taken later on.
The $M\times N$ coupling matrix $W$ has elements
\begin{equation}
\nonumber
W_{mn} = \delta_{mn}\left(\frac{2M\delta}{\pi^2}\right)^{1/2}\!\!\!\!\!,\,
m=1,2,\ldots M,\, n=1,2,\ldots N.
\end{equation}
The energy $\delta$ is one half of the mean level spacing of $H$.
For the chaotic billiard, we assume that the Hermitian matrix $H$ has the
Pandey-Mehta distribution \cite{Mehta,PandeyMehta}
\begin{eqnarray}
& & P(H)\propto \nonumber\\
& & \exp\left(-\frac{M(1+\alpha^2)}{4 \lambda^2}
\sum_{i,j=1}^M\left[(\mbox{Re}H_{ij})^2+\alpha^{-2}
(\mbox{Im} H_{ij})^2\right]\right),
\label{eq:PM}
\end{eqnarray}
with $\lambda=2M\delta/\pi$. The parameter $\alpha \in [0,1]$ measures
the strength of the time-reversal symmetry breaking.
The relation between $\alpha$ and the magnetic flux $\Phi$ through a
two-dimensional billiard (area $A$, no impurities, Fermi velocity
$v_{\rm F}$) is \cite{Bohigas95,FrahmPichard95}
\begin{equation}
\label{eq:crossoverpar}
M\alpha^2=c (\Phi e/h)^2 \hbar v_{\rm F} (A \delta^2)^{-1/2},
\end{equation}
with $c$ a numerical coefficient of order unity.
(For example, $c=\frac{2}{3}\sqrt{\pi}$ for a circular billiard which is
chaotic because of diffuse boundary scattering \cite{FrahmPichard95}.)
Time-reversal symmetry is effectively broken when $M\alpha^2 \simeq N$,
which occurs for $\Phi \ll h/e$.
The effect of such weak magnetic fields on the bulk superconductor can
be ignored.

A key step \cite{Frahmetal} is to write the determinantal equation
(\ref{eq:det}) as an eigenvalue equation for an effective Hamiltonian
$H_{\rm eff}$,
\begin{mathletters}
\begin{eqnarray}
  & & \mbox{Det}(E-H_{\rm eff})=0,\\
  & & H_{\rm eff} = \left(\begin{array}{cc}
    H & -\pi W e^{i\mbox{\boldmath $\scriptstyle \phi$}}W^{\rm T} \\
    - \pi W e^{-i\mbox{\boldmath $\scriptstyle \phi$}}W^{\rm T} & -H^*
    \end{array}\right).\label{eq:HBdG}
\end{eqnarray}
\end{mathletters}%
The average density of states $\rho(E)$ of the chaotic billiard results from
\begin{equation}
\rho(E) = -\pi^{-1} \mbox{Im}\, \mbox{Tr}\, {\cal G}(E + i0^+),\quad
{\cal G}(z)=\langle(z-H_{\rm eff})^{-1}\rangle,
\label{eq:resolvent}
\end{equation}
where $\langle \cdots \rangle$ denotes an average of $H$ with distribution
(\ref{eq:PM}).
The matrix Green function ${\cal G}$
inherits the $2 \times 2$ block structure from the effective Hamiltonian
$H_{\rm eff}$. Because of this block structure, it is convenient to define
the $2 \times 2$ Green function
\begin{equation}
  {G} = \left(\begin{array}{cc} G_{11} & G_{12}\\G_{21} & G_{22}\end{array}
 \right) = \frac{\lambda}{M}
    \left( \begin{array}{cc} \mbox{Tr}\, {\cal G}_{11} & \mbox{Tr}\, {\cal
G}_{12} \\
           \mbox{Tr}\, {\cal G}_{21} & \mbox{Tr}\, {\cal G}_{22} \end{array}
\right).
\end{equation}
Using the diagrammatic method of Refs.\ \onlinecite{Pandey} and
\onlinecite{BrezinZee},
adapted
for a matrix Green function, we have derived a self-consistency equation for
$G$. To highest order in $1/M$, this equation reads
\begin{equation}
\label{eq:pastur}
{G} = \frac{\lambda}{M}\sum_{n=1}^M \left(
\begin{array}{cc}
z-\lambda {G}_{11} & \pi w_n^2 +x\lambda {G}_{12} \\
\pi w_n^{*2}+x\lambda {G}_{21} & z-\lambda {G}_{22}
\end{array}\right)^{-1},
\end{equation}
where we have abbreviated $w_n^2=(W e^{i\mbox{\boldmath $\scriptstyle
\phi$}}W^{\rm T})_{nn}$
and $x=(1-\alpha^2) / (1 + \alpha^2)$.
Eq.\ (\ref{eq:pastur})
is complemented with the boundary condition
$G \to \lambda/z$ for $|z|\to \infty$.

We take the limit $M\to \infty$ at fixed $M\alpha^2$ and $\delta$, and
assume in addition that $N\gg 1$.
Eq.\ (\ref{eq:pastur}) then simplifies to
\begin{mathletters}
\label{eq:dysongamma}
\begin{eqnarray}
{G}_{11} & = & [\case{1}{2}(\Phi/\Phi_{\rm c})^2 G_{11}- \pi z/N \delta]\times
\nonumber\\
& &[G_{12}^2 +G_{12}/\cos(\phi/2)],\\
{G}_{22} & = & {G}_{11},\ {G}_{21}={G}_{12},\ {G}_{12}^2 = 1+{G}_{11}^2,
\end{eqnarray}
\end{mathletters}%
where we have defined the critical flux $\Phi_{\rm c}$ by
\begin{equation}
\label{eq:critical}
M\alpha^2=\case{1}{8}N(\Phi/\Phi_{\rm c})^2.
\end{equation}
The solution of Eq.\ (\ref{eq:dysongamma}) for $\phi=0$ and $\phi=5\pi/6$
is plotted in Fig.\ \ref{fig:fig1} for several values of $\Phi/\Phi_{\rm c}$.
For $\Phi=0$ and $\phi=0$ the excitation gap equals $E_{\rm gap}=a E_{\rm T}$,
with $E_{\rm T}=N\delta/2\pi$ and $a=2^{-3/2}(\sqrt{5}-1)^{5/2}\approx 0.6$.
The gap decreases with increasing flux $\Phi$ or phase difference $\phi$.
When $\phi=0$, the gap closes at the critical flux
$\Phi_{\rm c}\simeq (h/e)(N\delta/\hbar v_{\rm F})^{1/2}A^{1/4}$.
When $\phi=\pi$, there is no gap at any magnetic field.
For $\phi$ between $0$ and $\pi$, the gap closes at the flux
$\Phi_{\rm c}(\phi)$ given by
\begin{equation}
\Phi_{\rm c}(\phi)=
   \left[\frac{2\cos(\phi/2)}{1+\cos(\phi/2)}\right]^{1/2}\Phi_{\rm c}.
\end{equation}

\begin{figure}
\epsfxsize=0.95\hsize
\epsffile{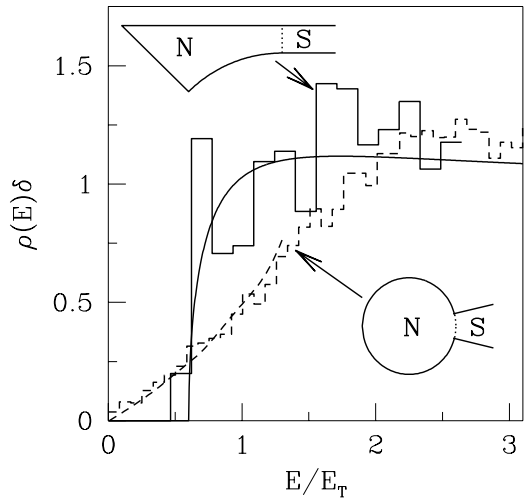}
\caption{\label{fig:fig2}
Histograms: density of states of a billiard coupled to a superconductor,
computed from Eq.\ (\ref{eq:resdet}) and averaged over a range of
Fermi energies. A chaotic Sinai billiard (top inset, solid histogram) is
contrasted with an integrable circular billiard (bottom inset, dashed
histogram). The number of propagating transverse modes at the
normal-metal---superconductor interface (dotted line in the insets) equals
$N=20$ in the chaotic billiard and $N=30$ in the circular billiard.
The solid curve is the prediction from random-matrix theory, the dashed curve
is the prediction from the Bohr-Sommerfeld approximation.
}
\end{figure}

So far we have used random-matrix theory to describe the chaotic system.
As a test, we can compute the exact quantum mechanical density of states of
a specific billiard, coupled to a superconductor. Following Doron, Smilansky,
and Frenkel \cite{Doronetal} we study a segment of a Sinai billiard,
drawn to scale in the top inset of Fig.\ \ref{fig:fig2}.
The scattering matrix $S_0(E)$ is determined by
matching wave functions at the dotted line separating the billiard (area $A$)
from the lead (width $W$). The NS interface is also chosen at the dotted
line. The density of states $\rho(E)$ follows
from Eq.\ (\ref{eq:resdet}). We average $\rho(E)$ over a small variation of
$E_{\rm F}$, such that the number of modes
$N=\mbox{Int}[mv_{\rm F}W/\pi\hbar]$ in the lead does not change.
The result for $N=20$ is shown in Fig.\
\ref{fig:fig2} (solid histogram), and is seen to agree quite well with the
prediction from random-matrix theory (solid curve). There is no adjustable
parameter in this comparison, the mean level spacing $\delta$ following
directly from $\delta=\pi \hbar^2/mA$.

We now turn from a chaotic to an integrable  billiard.
In Ref.\ \onlinecite{Melsenetal}, we computed the density of states of
a rectangular billiard, coupled to a superconductor by a narrow lead,
and found a linearly vanishing $\rho(E)$ for small $E$. Here,
we present data for a circular billiard (radius $R$)
in support of our claim that the absence of an excitation gap is generic for
integrable billiards. The circular billiard considered is drawn to scale in
the bottom inset of Fig.\ \ref{fig:fig2}. The scattering matrix $S_0(E)$ is
again determined by matching wave functions at the dotted line,
which also determines the location of the NS interface. A wedge-shaped
lead (opening angle $\theta$) is chosen in order not to break the rotational
symmetry (which simplifies the calculations).
The density of states is averaged over a range of Fermi energies at fixed
$N=\mbox{Int}[mv_{\rm F}R \theta /\pi\hbar]$. The result for $N=30$ is the
dashed histogram in Fig.\ \ref{fig:fig2}.

The density of states in the integrable billiard can be approximated
with the Bohr-Sommerfeld quantization rule \cite{Melsenetal},
\begin{equation}\label{eq:BS}
\rho_{\rm BS}(E) = N\int_0^{\infty}\!\!\! d s\, P(s)
\sum_{n=0}^{\infty}\delta\left(E-(n+\case{1}{2})\pi \hbar v_{\rm F}/s\right).
\end{equation}
Here $P(s)$ is the classical probability that an electron entering the
billiard will exit after a path length $s$.  We have calculated $P(s)$ for
the circular billiard shown in Fig.\ \ref{fig:fig2}, by generating a
large number of classical trajectories.
The resulting density of states $\rho_{\rm BS}(E)$ is in good agreement
with the quantum mechanical result in Fig.\ \ref{fig:fig2}.

To conclude, we have calculated the density of states of a chaotic Sinai
billiard, connected to a superconductor. The result is in good agreement
with the prediction from random-matrix theory \cite{Melsenetal}.
The excitation gap closes at a critical flux $\Phi_{\rm c}$ through the
billiard. In order of magnitude,
$\Phi_{\rm c} \simeq (h/e)(\tau_{\rm ergodic}/\tau_{\rm dwell})^{1/2}$,
where $\tau_{\rm dwell}$ is the mean dwell time of an electron in the
billiard and $\tau_{\rm ergodic}$ is the time required to explore the
entire available phase space. The precise value of $\Phi_{\rm c}$ depends
on the shape of the billiard, but the dependence on the tunnel
probability $\Gamma$ and the phase difference $\phi$ is universal:
\begin{equation}
\Phi_{\rm c}(\Gamma,\phi)= \Phi_{\rm c}(1,0)
\left[\frac{2\cos(\phi/2)}{\cos(\phi/2)-1+2/\Gamma}\right]^{1/2}.
\end{equation}
We have shown that the excitation gap, generic for chaotic billiards,
is absent in an integrable circular billiard. Instead, the density of
states $\rho(E)\propto E$ for small $E$, as in the rectangular billiard
considered previously \cite{Melsenetal}. The agreement with the
semi-classical Bohr-Sommerfeld approximation is better for the circular
than for the rectangular billiard. A rigorous semi-classical theory
for this problem remains to be developed.

This work was supported by the Dutch Science Foundation NWO/FOM and by the
Human Capital and Mobility program of the European Community.

\end{document}